\documentclass[10pt,preprint2,a4paper]{aastex}
\usepackage[left=2.1cm,top=2cm,right=1cm,nohead,nofoot]{geometry}
\usepackage{graphics,epsf}
\usepackage{amsmath}                % American Mathematical Society package
\usepackage{amsfonts}               % American Mathematical Society fonts
\usepackage{amssymb}                % American Mathematical Society symbol
\usepackage{epsfig}                 % EPS figures
\def \s{~\rm{s}}
\def \km{~\rm{km}}

\def \AU{~\rm{AU}}
\def \erg{~\rm{erg}}

\def \yr{~\rm{yr}}

\def \days{~\rm{days}}

\begin{document}

\title{EXPLAINING THE EARLY EXIT OF ETA CARINAE FROM ITS 2009
X-RAY MINIMUM WITH THE ACCRETION MODEL}

\author{Amit Kashi\altaffilmark{1} and Noam Soker\altaffilmark{1}}

\altaffiltext{1}{Department of Physics, Technion $-$ Israel Institute of
Technology, Haifa 32000 Israel; kashia@physics.technion.ac.il;
soker@physics.technion.ac.il.}
%\setlength{\columnsep}{1cm}
%\small

\begin{abstract}
We use the accretion model to explain the early exit of $\eta$ Car
from its 2009 X-ray minimum.
In the accretion model the secondary star accretes mass from the primary wind
near periastron passage, a process that suppresses the secondary wind.
As the shocked secondary wind is responsible for most of the X-ray emission,
the accretion process accounts for the X-ray minimum.
The early exit from the 2009 X-ray minimum after four weeks, instead of
ten weeks as in the two previous minima, is attributed to the primary wind that during
the last minimum was somewhat faster and of lower mass loss rate than during the two
previous X-ray minima.
This results in a much lower mass accretion rate during the X-ray minimum.
We show that using fluctuations in these quantities that are within the range
deduced from fluctuations in the X-ray flux outside the minimum, can account
for the short duration of the last X-ray minimum.
The shorter X-ray minimum may have further implications on the recovery
of the system from the spectroscopic event.
\end{abstract}


% ==========================================================
\section{INTRODUCTION}
\label{sec:intro}
% ==========================================================

For $2023\days$ we have been waiting for the intriguing periastron
passage of the massive binary system $\eta$ Car.
And indeed, we were not disappointed, as the new X-ray observations (Corcoran 2009)
brought some surprises.
The periodicity itself, which is observed in all wavelengths (e.g., radio, Duncan \& White
2003; IR, Whitelock et al. 2004; visible, van Genderen et al. 2006,
Fernandez Lajus et al. 2008;
UV, Smith et al. 2004; emission and absorption lines, Nielsen et al. 2007
Damineli et al. 1997, 2008a, b; X-ray, Corcoran 2005, 2008, Hamaguchi et al. 2007),
is attributed to a binary orbital period (Damineli 1996).

The X-ray light curve was followed  by RXTE (Corcoran 2009) during three minima.
In all three minima the X-ray intensity increases prior to the start of
the X-ray minimum and then drops sharply to minimum, which was about the same
level in the three minima (Corcoran 2005, 2009).
However, while the 1998 and 2003.5 minima lasted $65-70\days$ (Corcoran 2005),
the exit from the last X-ray minimum started only $27\days$ after periastron
passage (Corcoran 2009); last periastron passage is taken at January 11, 2009.

In Kashi \& Soker (2009a; see also 2009b) we have already pointed out the
possibility that the duration of the X-ray minimum can be shorter or longer.
We considered relatively small variations in the primary wind properties,
and estimated that the X-ray minimum can be different by several days.
In this {\it Letter} we reexamine X-ray variations far from the minimum, i.e., near apastron,
and deduce that the variations (fluctuations) in the primary wind velocity and mass
loss rate can be up to a factor of two (\S \ref{sec:xray}).
We then show (\S \ref{sec:accrate}) that within the \emph{accretion model}
(Soker 2005; Akashi et al. 2006; Kashi \& Soker 2008, 2009a) such an increase
in the primary wind velocity near periastron passage can account for the early
exit of $\eta$ Car from the X-ray minimum.
In the accretion model the X-ray minimum is attributed to the accretion of the
primary wind by the secondary star.
The accretion process suppresses the secondary wind, that otherwise is responsible for most
of the X-ray emission (Corcoran et al. 2001; Pittard \& Corcoran 2002; Akashi et al. 2006;
Henley et al. 2008; Okazaki et al. 2008; Parkin et al. 2009).
Our discussion and summary is in \S \ref{sec:summary}.

The present study presents neither new ingredients to the accretion model nor new
type of calculations. The essence of this paper is to strengthen the accretion model
by showing that it can easily accommodate the new observations by only changing the
variables that are most likely to change from orbit to orbit and along the orbit,
namely, the primary wind properties.

% ==========================================================
\section{POSSIBLE FLUCTUATIONS IN THE PRIMARY WIND }
\label{sec:xray}
% ==========================================================

The binary parameters are as in our previous paper
(Kashi \& Soker 2009a, where references are given):
The assumed stellar masses are $M_1=120 \rm{M_\odot}$ and $M_2=30 \rm{M_\odot}$,
the eccentricity is $e=0.9$, and the orbital period is $P=2024 \days$.
The stellar winds' mass loss rates and terminal velocities that were used are
$\dot M_1=3 \times 10^{-4} \rm{M_\odot\yr^{-1}}$, $\dot M_2 =10^{-5} \rm{M_\odot\yr^{-1}}$,
$v_{\rm 1,\infty}=500 \km \s^{-1}$ and $v_{\rm 2,\infty}=3000 \km \s^{-1}$.
While in our previous papers the mass loss rate and terminal velocity of the primary wind
were held constant with the above values, in the present paper we vary them to
account for the early exit from the X-ray minimum.

To estimate the largest fluctuations in the primary wind properties,
we use the fluctuations in the X-ray luminosity near apastron in the following manner.
The hard X-ray emission observed in $\eta$ Car is emitted by the shocked secondary wind
(Corcoran et al. 2001; Pittard \& Corcoran 2002; Akashi et al. 2006; Henley et al. 2008;
Okazaki et al. 2008; Parkin et al. 2009).
For constant secondary wind properties the X-ray luminosity varies as $\sim D_2^{-1}$,
where $D_2$ is the distance of the stagnation point of the colliding winds from the
secondary star.
This can be understood from two different considerations.
When the radiative cooling time is long, the X-ray emission is (Akashi et al. 2006)
$L_x \simeq 0.5 \dot M_{2s} v_2^2 (\tau_{f2}/\tau_{\rm cool2})$, where $\tau_{f2}\simeq D_2/v_2$
is the outflow time of the shocked gas, and $\tau_{\rm cool2} \propto n^{-1}$ is the
radiative cooling time of the shocked secondary wind.
The postshock secondary wind density varies as $n \propto D_2^{-2}$.
We find therefore $L_x \propto D_2^{-1}$.
In the second approach the X-ray emissivity (power per unit volume) $\Lambda n^2$ is used.
The luminosity is $L_x =V_2 \Lambda n^2$, where $V_2 \propto D_2^3$ is the volume of the
shocked secondary wind. The mass in the volume $V_2$ is proportional to its
outflow time $\tau_{f2}\simeq D_2/v_2$. Therefore, $n \propto D_2^{-2}$ (as before), and
again we recover the relation $L_x \propto D_2^{-1}$.

The distance $D_2$ is given by equating the momentum fluxes of the two winds.
When $\eta$ Car is near apastron (hence orbital velocity is negligible)
$D_2$ is approximately given by $D_2 = r \zeta / (1 +\zeta)$, where here we define
$\zeta \equiv (\dot M_2 v_2/\dot M_1 v_1)^{1/2}$, and $r$ is the orbital separation.
For our typical parameters $\zeta=0.45$.
Using this relation in the expression for the X-ray luminosity gives
\begin{equation}
L_x = K(r,v_2,\dot M_2) \frac{1 +\zeta}{\zeta},
\label{lx1}
\end{equation}
where the function $K$ depends on the orbital separation and secondary wind properties.
The logarithmic variation of the X-ray luminosity with respect to the primary wind
momentum discharge $\dot p_1 \equiv \dot M_1 v_1$ is given by
\begin{equation}
\frac{dL_x}{L_x} = \frac{1}{2(1+\zeta)} \frac {d \dot p_1}{\dot p_1}.
\label{dlx1}
\end{equation}

{}From the X-ray light curve (Corcoran 2005) we find that the variations between
cycles and during a short time within one cycle can be ${dL_x}/{L_x} \simeq \pm 0.25$.
If we attribute this variation to changes in the primary wind properties, then we find
from equation (\ref{dlx1}) and $\zeta=0.45$ that ${d \dot p_1}/{\dot p_1}\simeq \pm 0.7$.
Basically, if the mass loss rate does not change the terminal velocity can vary in
the range $\sim 300-850 \km \s^{-1}$; such changes are seen in the solar wind.
If the wind velocity is larger when the mass loss rate is lower, then the
variation in the wind velocity can be even larger.
In the acceleration zone close to the primary star the variations can be much larger.
In such an eccentric, asynchronous binary system like $\eta$ Car,
tidal interaction between the stars can result in kinetic energy dissipation
through viscous shear.
This can lead to variations in mass-loss rate and wind velocity structure,
or even asymmetric mass-shedding (Koenigsberger \& Moreno, 2009).

Our conclusion is that variations in the primary wind velocity by
a factor of $\la 2$ in the acceleration region are reasonable. As
mentioned earlier, in our previous paper (Kashi \& Soker 2009a) we
assumed variations in $\dot p_1$ of no more than $\sim 10\%$, and
predicted the possibility of an early (or late) exit by several days.
Motivated by the new observations (Corcoran 2009), we
reexamined the primary wind variations, and concluded in this
section that we can allow for much larger variations.

% ==========================================================
\section{VARYING THE ACCRETION RATE AT THE LATEST PERIASTRON PASSAGE}
\label{sec:accrate}
% =====================================================

In Kashi \& Soker (2009a) we took two extreme processes, that we expect to bound the true
mass and angular momentum accretion rates, to estimate the mass and angular momentum accretion rate.
These are the Bondi-Hoyle-Lyttleton (BHL) accretion process from a wind,
and a Roche lobe overflow (RLOF) type mass transfer.
Very close to periastron passage, $\vert t \vert \la 10~$day,
the accretion process is an hybrid of the BHL and the
RLOF mass transfer processes, but at the end of the accretion phase the
accretion will be of the BHL type.
The calculated accreted mass in the different models considered was estimated to be
in the range $M_{\rm acc} \sim 0.4 - 3.3 \times10^{-6} \rm{M_\odot}$,
and the accretion rate was typically
$\dot{M}_{\rm acc} = 5 \times10^{-7} - 5 \times10^{-5}  \rm{M_\odot\yr^{-1}}$,
with the higher values close to periastron.

The accretion processes close to periastron is very complicated, and
for its accurate study one must use 3D hydrodynamical numerical codes.
In this highly eccentric binary system the primary angular velocity is expected
to be much below the angular velocity of the secondary near periastron.
Therefore, the mass transfer process will not be as in the RLOF
process in synchronized binary systems.
For that reason, in the present paper we use only the BHL accretion process.

We take the $\beta$-profile to describe the primary wind acceleration
\begin{equation}
v_1(r_1)=v_s+(v_{\rm 1,\infty}-v_s)\left(1-\frac{R_1}{r_1}\right)^{\beta} ,
\label{eq:v1}
\end{equation}
where $R_1$ is the primary radius, $r_1$ distance from the primary center,
and $v_s=20  \km \s^{-1}$.
For our ``standard case'' we use the parameters $\beta=3$ and a
terminal wind velocity of $v_{\rm 1,\infty}=500 \km \s^{-1}$.
The orbital eccentricity is taken to be $e=0.9$.

\begin{figure}[!t]%[!ht]
\resizebox{0.50\textwidth}{!}{\includegraphics{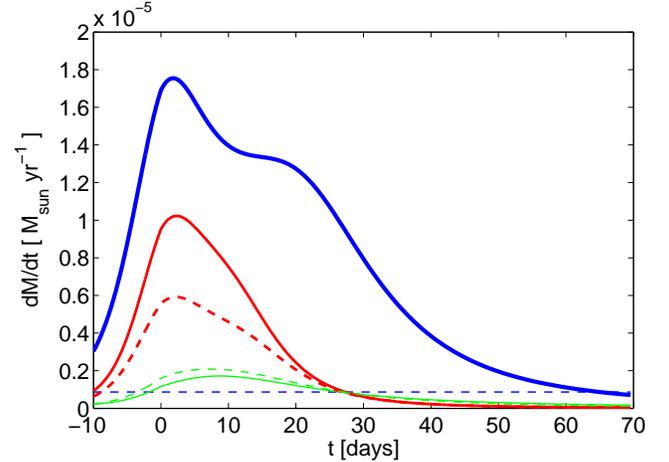}}
%\vskip -9.2 cm
%\hskip -3.5 cm  % FOR ONE PAGE
\caption{\footnotesize The variation of accretion rate near periastron passage.
Only the Bondi-Hoyle-Lyttleton (BHL) accretion prescription is presented.
In the first $\sim 20$ days extra accretion is possible via a RLOF type mass transfer.
The different lines, from top to bottom, depict the mass accretion rate
for the following cases characterized by primary mass loss rate in units of
$M_\odot \yr^{-1}$, the value of $\beta$ in the acceleration profile given by
equation (\ref{eq:v1}), and the primary wind terminal velocity give in units of $\km \s^{-1}$,
$(\dot M_1, \beta, v_{\rm 1,\infty})$:
Solid very thick blue line: The standard case that was used in our previous paper $(3\times 10^{-4},3, 500)$;
Solid thick red line: $(3\times 10^{-4},3,900)$;
Dashed thick red line: $(1.5\times 10^{-4},3,780)$;
Dashed thin green line: $(1.5\times 10^{-4}, 1.05,500)$;
Solid thin green line: $(3\times 10^{-4}, 0.35, 500)$.
The dashed thin blue line is the reference accretion rate as given by equation (\ref{dmacc65}).
}
\label{fig:accrate}
\end{figure}

The mass accretion rate for the standard case was calculated in our previous paper
(Kashi \& Soker 2009a).
The BHL mass accretion rate as calculated there is drawn by the
solid very thick blue line in Figure \ref{fig:accrate}.
At $t=65\days$, where the system starts leaving the 1998 and 2003.5
X-ray minima, the mass accretion rate is
\begin{equation}
\begin{split}
\dot M_{\rm ref} \equiv \dot{M}_{\rm acc}(t=65\days,\beta=3,v_{\rm 1,\infty}=500 \km \s^{-1}) = \\
8.7 \times10^{-7} \rm{M_\odot\yr^{-1}}.
\end{split}
\label{dmacc65}
\end{equation}
This is depicted by the dashed horizontal thin blue line in Figure
\ref{fig:accrate}. We take this value to be the reference value
for the accretion rate at which the secondary wind rebuilds itself
after the accretion phase. We note that for the present goal the
exact value of $\dot M_{\rm ref}$ is not important. What is
important is how we vary the primary wind properties to obtain the
same value at a much early time.

We emphasize that in our model for an early exit relative to the previous two cycles,
it is required that the primary wind velocity in the last
minimum was higher than the wind velocity in the previous two minima.
It is not required that the primary wind velocity after periastron be higher
than the primary wind velocity far from periastron.
In the present paper we take the velocity of the primary wind in the previous two minima
to be $v_1= 500 \km \s^{-1}$.

We now search for the value $v_{\rm 1,\infty New}$ in the $\beta=3$ profile
that would give an accretion rate of $8.7 \times10^{-7}  \rm{M_\odot\yr^{-1}}$
at $t=27\days$, the time when the early exit from the 2009 X-ray minimum started.
We keep the mass loss rate at $\dot M_1 =  3 \times 10^{-4} \rm{M_\odot\yr^{-1}}$.
We find that $v_{\rm 1,\infty New}=900 \km \s^{-1}$ gives the desired accretion
rate, as shown by the solid thick red line in Figure \ref{fig:accrate}.
This is within the reasonable range of the primary wind velocity fluctuations
we estimated in the previous section.

We also find that if we keep $v_{\rm 1,\infty}=500 \km \s^{-1}$
and $\dot M_1 =  3 \times 10^{-4} \rm{M_\odot\yr^{-1}}$ as in the standard case,
then a value of $\beta=0.35$ gives the desired accretion rate at $t=27 \days$.
This case is presented by the solid thin green line in Figure \ref{fig:accrate}.

It is likely that a more efficient acceleration process occurs when the mass loss rate is lower.
We assume now that the primary mass loss rate was lower in the 2009 minimum, which lead to
a more efficient acceleration.
To demonstrate the feasibility of our model, we take the mass loss rate to have been half
its typical value $\dot M_{1,\rm{New}} =  1.5 \times 10^{-4} \rm{M_\odot\yr^{-1}}$.
For $\dot M_{1,\rm{New}}$ and keeping $\beta=3$, we find that $v_{\rm 1,\infty New2}=780 \km \s^{-1}$
gives the reference accretion rate $\dot M_{\rm ref}$ at $t=27\days$
(dashed thick red line in Figure \ref{fig:accrate}).
For $\dot M_{1,\rm{New}}$ and keeping $v_{\rm 1,\infty}=500 \km \s^{-1}$, we find that $\beta=1.05$
gives the reference accretion rate at $t=27\days$ (dashed thin green line in Figure \ref{fig:accrate}).
%%%%%%%%%%%%%%%%%%%%%%%%%%%%%%%%%%%%%%%%%%%%%%%%%%%%%%%%%%%%%%%%%%%%%%%%%
%%%%%%%%%%%%%%%%%%%%%%%%%%%%%%%%%%%%%%%%%%%%%%%%%%%%%%%%%%%%%%%%%%%%%%%%%

As well, our model might account for the change in the rate of the
exit from the two events.
We quantify the observed exit rate by defining the average value of
$\Delta L_x/\Delta t$ over the exit from minimum time period.
Here $\Delta t$ is the exit time interval between the last epoch in which the
X-ray luminosity has been at its minimum value and the first epoch in which the
X-ray luminosity has returned to its quiescent level,
and $\Delta L_x$ is the difference in the corresponding X-ray luminosities.
Using the RXTE results (Corcoran 2005; 2009), we find that
the (averaged) observed exit rate of the 2003.5 minimum was
$\Delta L_x/\Delta t = 1.92 \times 10^{28} \erg s^{-2}$.
The observed exit rate of the 1998 was lower by $11\%$, while
the observed exit rate of the 2009 minimum was substantially lower,
with $\Delta L_x/ \Delta t=1.39 \times 10^{28} \erg s^{-2}$,
which is $~72 \%$ of the observed exit rate of the 2003.5 minimum.

We check whether the change in time evolution of the mass accretion rate during
the exist from the X-ray minimum might explain the change in the observed exist rate.
For that, we measured the average rate of change in the mass accretion rate,
$\Delta \dot M_{\rm acc}/\Delta t$, as given in Figure \ref{fig:accrate}.
The average is over the exit time interval of each event.
We summarize the results in Table \ref{Table:compare}.
We find that the ratio in the average values ($\Delta \dot M_{\rm acc}/\Delta t$) during the
2009 and the 2003.5  minima is 0.68.
The proximity of the ratio 0.68 (ratio of modelled accretion rate in the two minima)
to 0.72 (ratio of observed X-ray flux time evolution in the two minima) should not
be given a strong weight, as we do not expect a simple relation between mass accretion
rate and X-ray luminosity.
What is important is the general qualitative change.
We conclude that our model can account, with the same change in
parameters, to both the 2009 early exit from the minimum,
and for the 2009 slower rate of increase in the X-ray flux during the exit from minimum.
\begin{table}[!t]
\begin{tabular}{||l||c|c||}
\hline \hline
Cycle           &2003.5               &2009\\
\hline
\multicolumn{3}{||c||}{Observations}\\
\hline
$t_1$ [days after periastron]       &64.7                 &26.3\\
$t_2$ [days after periastron]       &91.0                 &64.7\\
$L_{x1}$ [$\erg \s^{-1}$]              &$1.9\times10^{34}$  &$1.9\times10^{34}$\\
$L_{x2}$ [$\erg \s^{-1}$]              &$6.3\times10^{34}$  &$6.5\times10^{34}$\\
$\Delta L_x/\Delta t$ [$\erg \s^{-2}$]              &$1.92\times10^{28}$ &$1.39\times10^{28}$\\
Ratio (2003.5/2009)                 &\multicolumn{2}{|c||}{0.72}\\
\hline
\multicolumn{3}{||c||}{Model}\\
\hline
$(\dot M_{\rm acc})_1$ [$\rm{M_\odot\yr^{-1}}$]    &$10^{-6}$   &$10^{-6}$\\
$(\dot M_{\rm acc})_2$ [$\rm{M_\odot\yr^{-1}}$]    &$0$                  &$0$\\
$\Delta \dot M_{\rm acc}/\Delta t$ [$\rm{M_\odot\yr^{-2}}$]  &$-1.4\times10^{-5}$  &$-9.5\times10^{-6}$\\
Ratio (2003.5/2009)                     &\multicolumn{2}{|c||}{0.68}\\
\hline \hline
\end{tabular}
\caption{\footnotesize Comparing the observed exit rates ratio to the one obtained from our model,
if we assume that the X-ray exit rate is proportional to the change in
the mass accretion rate (although in reality no simple relation is expected).}
\label{Table:compare}
\end{table}

% ==========================================================
\section{DISCUSSION AND SUMMARY}
\label{sec:summary}
% =====================================================

Our aim was to explain the early exit of $\eta$ Car from its X-ray minimum in the last cycle.
Whereas in the previous 2 cycles the exit occurred $\sim 10$~weeks
after periastron passage, in the last cycle it occurred only $\sim 4$~weeks after
periastron passage.
It was known that in the accretion model variations in the primary wind mass loss rate and/or
velocity can cause variations in the duration of the X-ray minimum (Kashi \& Soker 2009a).
However, the exit was earlier than anticipated in our earlier paper.
The reason is that in our previous paper we considered relatively small
variations in the primary wind properties.

In the accretion model the secondary star accretes mass from the primary wind near periastron.
While for most of the $\sim 5.54\yr$ orbital period the secondary gravity has a negligible effect
on the colliding winds cone, near periastron the shocked primary wind is very close to the
secondary star and very dense, and the secondary gravitational field becomes non-negligible.
As was shown in previous papers (Soker 2005; Akashi et al. 2006; Kashi \& Soker 2008a, 2009a) the
gravitational field of the secondary ensures accretion.
The accretion process is assumed to shut down the secondary wind, hence the main X-ray source;
the system then enters the X-ray minimum.

Although $\eta$ Car had an early exit from the X-ray minimum in the last cycle,
the beginning of the minimum was as in the previous two cycles, and occurred shortly
before phase zero.
The onset of the accretion phase requires an almost extinction of the secondary wind,
and therefore a high mass accretion rate that occurs just before periastron.
During the phase interval $\sim 0.01$ to $-0.01$  ($-20 \la t \la \sim 20~$days)
the two stars are very close ($r \la 3 \AU)$, and the mass accretion rate is an hybrid of
a RLOF and an accretion from a wind, with the RLOF type most likely the dominant
process very close to periastron passage (Kashi \& Soker 2009a; after $t \simeq 20~$days
the accretion from the wind is the relevant process, and for that it was used in the paper).
Therefore, the primary wind velocity does not influence much the mass accretion
rate during the onset of the X-ray minimum.

The above explanation holds for small and moderate fluctuations in the
primary wind properties.
The very early exit from the X-ray minimum in the last cycle suggest that the changes
in the primary wind properties were large.
Therefore, for our model to work we require than the changes occurred near periastron passage,
most likely due to the strong tidal interaction.
Most likely, the tidal interaction amplified a small internal change in the wind
properties, e.g., as might results from magnetic activity.
Koenigsberger \& Moreno (2009) calculations show indeed that
tidal interaction can cause large changes in the wind properties of
LBV stars in binary systems.

Following recent RXTE observations by Corcoran (2009),
we reexamined the behavior of the X-ray emission when the system is far from
the X-ray minimum, i.e., it is near apastron.
Based on the behavior of the X-ray emission, in section \ref{sec:xray} we found
indeed that the variations (fluctuations) in the mass loss rate and/or velocity of the
primary wind can be by up to a factor of $2$.

Armed with this finding, we examined the possible implications for the accretion model.
We assumed that the secondary wind, and hence the X-ray emission, is recovered when the
accretion rate is as in $t=65\days$ (after periastron) in the previous cycles.
The case with the parameters used to explain the previous 2 cycles is termed ``the standard case''.
We found that by increasing the primary wind velocity by a factor $\le 2$,
and possibly with lowering the primary wind mass loss rate,
we can have the mass accretion rate at $t=27 \days$
as in $t=65 \days$ in the previous cycles.
We examined several cases, as drawn in Figure \ref{fig:accrate}.

It should be emphasized that we did not add any new ingredients to the accretion model,
nor did we change the type of calculations.  The new addition is simply repeating the
calculation from our previous papers with different primary wind properties.
As discussed, these properties are likely to vary by a large factor
from orbit to orbit and along the orbit.
This new part shows that the accretion model is quite robust.

One implication of our finding is that in some lines that come from the
primary wind during the minimum the radial velocity in the last minimum
was larger than in the previous minima.
We emphasize that what matters to the early exit from the minimum is
the primary wind velocity in the last cycle compared with the previous two minima,
and that the wind velocity in the minimum can be {\it lower}
than the average wind velocity far from the minimum.
We note also that for accretion the velocity of the wind expelled
in and near the equatorial plane matters.
The wind toward our line of sight can have somewhat different value.
We observe the binary system at inclination angle of $41^\circ$ (Smith 2006),
therefore such variations might not be as large, as tidal effects
are stronger in the equatorial plane."

The shorter X-ray minimum might have further implications.
As the secondary accreted less mass than in the previous two cycles,
the accretion disc (or belt) formed around it may be thinner,
and its dissipation time would be shorter.
Therefore the recovery time of different spectral lines and electromagnetic bands
from the minimum can change.

In the same way as the last X-ray minimum was shorter, future X-ray minima
might be longer than $70\days$.
The point is that any fluctuation in the wind, even a moderate one,
can change the duration of the X-ray minimum.

We thank Nathan Smith and an anonymous referee for helpful comments.
This research was supported by grants from the Israel Science Foundation,
and a grant from the Asher Space Research Institute at the Technion.

\end{document}